\documentclass[superscriptaddress,prl,longbibliography, amsmath,amssymb,reprint,showkeys,showpacs]{revtex4-1}

\usepackage{graphicx}
\usepackage{graphics}
\usepackage{amssymb}
\usepackage{amsmath,bm}
\usepackage{float}
\usepackage[usenames,dvipsnames]{color}
\usepackage[export]{adjustbox}
\usepackage{hyperref}
\usepackage{gensymb}
\usepackage{tipa}
\usepackage{appendix}

\usepackage{soul,xcolor}
\setstcolor{red}

\hypersetup{
    colorlinks=true,
    linkcolor=blue,
    citecolor=blue,
    filecolor=cyan,      
    urlcolor=cyan,
}
\newcommand{\ket}[1]{\left|  #1 \right\rangle}

\begin{document}

\title{Correlation Localization in Waveguide QED with Delayed Interactions}


\author{N. Vera}
\email{nicvera@udec.cl}
\affiliation{Departamento de F\'isica, Facultad de Ciencias F\'isicas y Matem\'aticas, Universidad de Concepci\'on, Concepci\'on, Chile}

\author{F. M. Quinteros}
\affiliation{Departamento de F\'isica, Facultad de Ciencias F\'isicas y Matem\'aticas, Universidad de Concepci\'on, Concepci\'on, Chile}

\author{P. Barberis-Blostein}
\affiliation{Instituto de Investigaciones en Matemáticas Aplicadas y en Sistemas, Universidad Nacional Autónoma de México, Ciudad Universitaria, 04510 DF, Mexico}

\author{P. Solano}
\email{psolano@udec.cl}
\affiliation{Departamento de F\'isica, Facultad de Ciencias F\'isicas y Matem\'aticas, Universidad de Concepci\'on, Concepci\'on, Chile}

\begin{abstract}
We study the atom-atom correlation length in an atomic array coupled to a waveguide under the Bragg condition with delayed non-Markovian interactions caused by a finite photon propagation time. Starting from a single excited atom, the excitation partially spreads among all atoms, reaching a steady state. The remaining excitation localizes near the initially excited atom, and the atom-atom correlation length decreases as a power law with the interaction delay. This localization phenomenon reveals how the delay-induced non-Markovian behavior affects the correlation transport in waveguide QED systems.
\end{abstract}
\maketitle

{\it Introduction.---} Waveguide quantum electrodynamics (wQED) describes several phenomena that emerge from atomic ensembles interacting with guided electromagnetic (EM) modes \cite{roy2017,chang2018, sheremet2023,Ciccarello_2024Wqed,arno_2010,Nieddu2016Jjop,solano2017optical, lechner2023}. The one-dimensional nature of waveguides enhances light-matter coupling and collective effects, such as superradiance, subradiance, and photon-mediated entanglement, which are not readily accessible in free-space configurations \cite{chang2012,zheng2013,Kimble_2015, Asenjo_2017,NatureCorzo2019}. By coupling arrays of artificial or real atoms to one-dimensional waveguides, it is possible to engineer strong tunable interactions mediated by the guided EM field \cite{schilke2011, Olmos_2021,Tecer_2026}, which has enabled breakthroughs in quantum information processing, quantum networks, and the exploration of many-body quantum phenomena \cite{zheng_2013qc,Paulisch_2016, pichler2017,ciraczoller1997, kimble_2008,Ramos_2016,douglas2015, cardenas2023many,Cirac_2025}.

A distinctive feature of wQED systems is their ability to enable long-range atom-atom interactions mediated by a guided EM field. When photon-mediated interactions are considered instantaneous owing to a negligible photon travel time compared to atomic timescales, the atomic state dynamics is well approximated by a Markovian evolution and simplifies to a memoryless process described by a Lindblad master equation \cite{Lehmberg1970,Lindblad1976, Carmichael1993}. However, as experimental platforms achieve larger interatomic separations, the propagation delay of photons becomes significant \cite{Wallraff_2013,Solano2017_inf,Wallraff_qLink_2020,ferreira2021,mohammad_2025,hughes_2026}. Under these conditions, the evolution of the atoms and EM modes becomes entangled, the Born approximation is no longer valid, and the atomic states exhibit non-Markovian dynamics \cite{Sinha_2020,Dominik_2024,Annyun_2025,Sinha_2025}. Such delayed interactions significantly modify collective atomic behaviors, giving rise to phenomena such as enhanced decay beyond the standard superradiance \cite{Sinha_2020} and bound states in the continuum (BIC) \cite{Calajo_2019}, predicting excitation storage in decoherence-free subspaces \cite{Alvarez_2024,Giuseppe_2025} and metrologically relevant configurations \cite{kanu_2026}, highlighting the potential for exploiting non-Markovianity to engineer quantum states.

Despite the many advances in delay-induced non-Markovian wQED \cite{Ramos_2016,pichler2016,carmele_2020,xinyou_2024,Sinha_2020,Dominik_2024,Annyun_2025,solano_2020_dist,dissimilar_2023,Sinha_2025,kanu_2026,maffei_2025,zueco_2021,guo_2020,Calajo_2019,Wallraff_2022,Alvarez_2024,ciccarello_2026,Giuseppe_2025}, most studies have focused on the propagation of atomic and photonic excitations, revealing how delayed interactions alter the collective decay and excitation dynamics of a system. However, the transport of correlations, which is key to understanding entanglement distribution, coherence preservation, and many-body quantum effects, remains largely unexplored in this context. A previous seminal study showed the effect of correlation localization under delayed interactions for chiral wQED systems during transient behavior \cite{Cirac_2025}. Nevertheless, the absence of back-action in chiral systems conceals fully collective effects and steady-state behaviors. These are key aspects to understand the correlation transport, which underpins phenomena such as Anderson-like and many-body localizations \cite{Anderson_1958,Abanin_2019}, where the interactions and disorder conspire to maintain localized quantum states over extended timescales. Therefore, examining how delay-induced non-Markovianity affects atomic correlation transport in collective systems, even within basic models containing a single excitation, is essential for advancing our understanding and control of quantum information in extended atomic arrays.

\begin{figure}
    \centering
    \includegraphics[width=\linewidth]{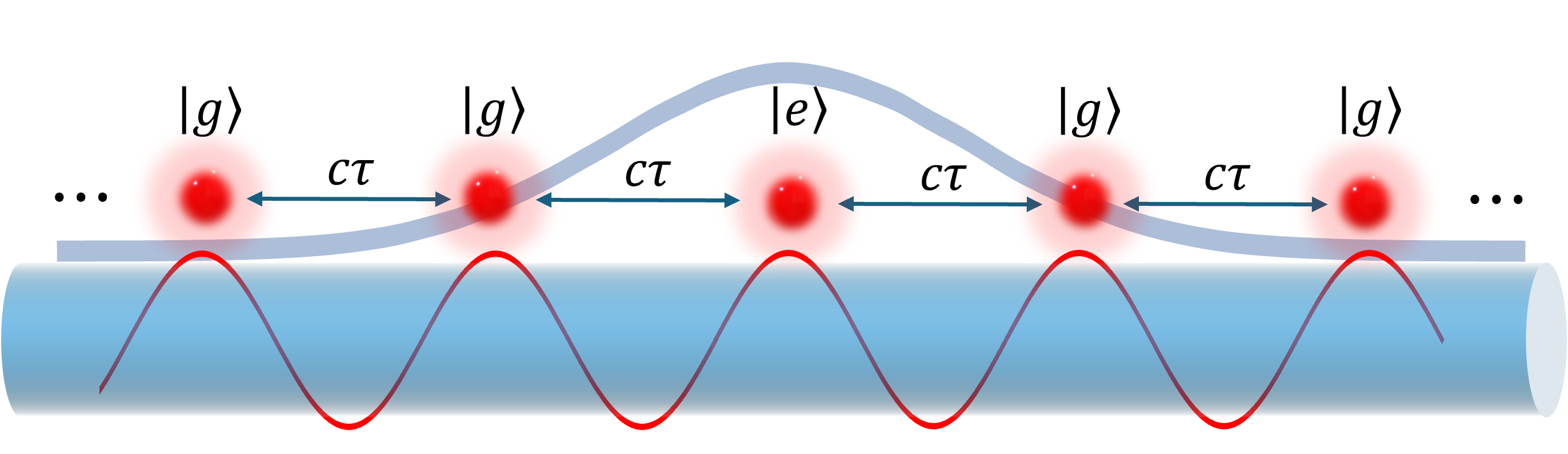}
    \caption{Schematic of an atomic array coupled to a waveguide under the Bragg condition, where all atoms are spaced by a distance $v \tau$, with $v$ representing the group velocity and $\tau$ the delayed interaction time. The central atom is initially excited, whereas all other atoms are in the ground state. The waveguide mediates photon exchange among the atoms, enabling collective dynamics that lead to localization of steady-state atom-atom correlations for large delays.}\label{fig:setup}
\end{figure}

In this paper, we explore how correlations localize in atomic arrays coupled to a waveguide under the Bragg condition and delay-induced non-Markovian interactions. We solve the dynamics of systems initialized with a single excited atom that evolves toward a long-time asymptotic collective steady state and study the atom-atom correlations through the array.  We compare the conventional Markovian regime, where correlations spread evenly across the array, with the delayed non-Markovian regime, where a finite photon travel time causes a partial excitation to remain near the initially excited atom. As the delay increases, the atom-atom correlation length decreases according to a power law, revealing a novel non-Markovian transport phenomenon. We also study how spatial disorder affects the localization of correlations, showing that delayed interactions further localize correlations over disorder, due to an interference mechanism specific to the Bragg condition. Our results elucidate novel scenarios for studying localization phenomena and controlling correlation confinement in wQED.

{\it Atomic dynamics in wQED with delay.---} We derive the atomic equations of motion using the Weisskopf-Wigner theory of spontaneous emission for $N$ atoms coupled to a waveguide in the single-excitation regime \cite{Weisskopf1930,PhysRevA.77.043833}, avoiding the Born-Markov approximation by following the discussion in \cite{milonni1995photodetection}. The interaction Hamiltonian of an ensemble of $N$ identical atoms coupled to a waveguide in the rotating wave approximation (RWA) and rotating frame is \cite{PhysRevA.91.042116}
{\small
\begin{multline}
    H_{\rm int}^{\rm wg} = i \hbar \sum_{\lambda = L, R}\sum_{j} \int d\omega \sqrt{\frac{\gamma_{\lambda}^{1D}}{2 \pi}}b_{\lambda}^\dagger (\omega)\sigma_j e^{i ( \omega x_j /v-(\omega-\omega_0)t) }\\
   + \rm H.c.,
\end{multline}
}

\noindent where $b_\lambda(\omega)$ ($b^{\dagger}_\lambda(\omega)$) is the bosonic annihilation (creation) operator of the EM field,  $\lambda = L,R$ denotes the right- and left-propagating modes, $v$ is the group velocity, $\gamma_\lambda^{1D}$ is the directional decay rate of a single atom into the waveguide, $\sigma_j$ ($\sigma^{\dagger}_j$) is the lowering (rising) operator of an excitation in the \textit{j}th atom, $x_j$ is the position of the \textit{j}th atom along the waveguide, and $\omega_0$ is the atomic resonance angular frequency.

We are interested in the relaxation dynamics that start from a single excited atom. This allows us to write the state of the system in the single-excitation manifold as
{\small
\begin{equation}
\ket{\psi}=\left(\sum_{j=1}^{N}\beta_j(t)\sigma^{\dagger}_j +\sum_{\lambda=L,R}\int d\omega \alpha_\lambda (t,\omega) b_{\lambda}^\dagger (\omega)\right)\ket{g_1 \dots g_N}\ket{0}
\end{equation}
}
Replacing into the Schrödinger equation and considering that the only decay channel is into the guided modes with symmetric probability ($\gamma_L=\gamma_R=\gamma$), one obtains the following set of differential equations for the evolution of the atomic coefficients (See Supplemental Material at \cite{supp} for a detailed derivation):
\begin{equation}\label{eq:mainresult}
    \dot{\beta}_j(t) = -\gamma\sum_{j'}e^{-i\omega_0\tau_{j,j'}}\beta_{j'}\left(t-\tau_{j,j'}\right)\Theta\left(t-\tau_{j,j'}\right)
\end{equation}
where $\tau_{j,j'}=\frac{|x_j-x_{j'}|}{v}$ is the delayed interaction time between the \textit{i}th and \textit{j}th atoms. We conveniently write this expression in a more concise vector form as 
\begin{equation}\label{eq:mainresultmat}
    \dot{\mathbf{B}}(t)=-\gamma \text{Diag}\left[\mathbf{\Gamma}\cdot\mathbf{B}(t-\tau_{j,j'})\right],  
\end{equation}
where $\mathbf{\Gamma}$ is the matrix of elements $\Gamma_{j,j'}=e^{-i\omega_0\tau_{j,j'}}$ and $\mathbf{B}(t-\tau_{j,j'})$ is the matrix of elements $B(t-\tau_{j,j'})=\beta_{j'}\left(t-\tau_{j,j'}\right)\Theta\left(t-\tau_{j,j'}\right)$. In this form, it is easy to solve the equations of motion using numerical methods. Notice that this equation is valid for any disposition of atoms along the waveguide and any initial condition in the atomic coefficients $\beta_j(t=0)$. For this study, we chose the interatomic distances such that adjacent atoms maintained a distance $d=c\tau$ (see Fig. (\ref{fig:setup})), and the propagation phase is a multiple of $2\pi$, which is known as Bragg condition, or mirror configuration \cite{chang2012,Olmos_2021,Sinha_2025}.

Although the Laplace transform can be used to solve the equations of motion, the analytical inversion of the system becomes intractable because of the intricate algebraic structure arising from the coupled differential equations with delay. These lead to transcendental characteristic equations in the Laplace space, involving multiple poles and non-rational functions, which grow combinatorially with the number of atoms, thereby exponentially increasing the complexity. Consequently, we employ numerical methods to solve delay differential equations, as they effectively handle time delays  directly in the time domain, bypassing the need for explicit inversion, while maintaining stability and accuracy in capturing the system dynamics.

\begin{figure}
    \centering
\includegraphics[width=\linewidth]{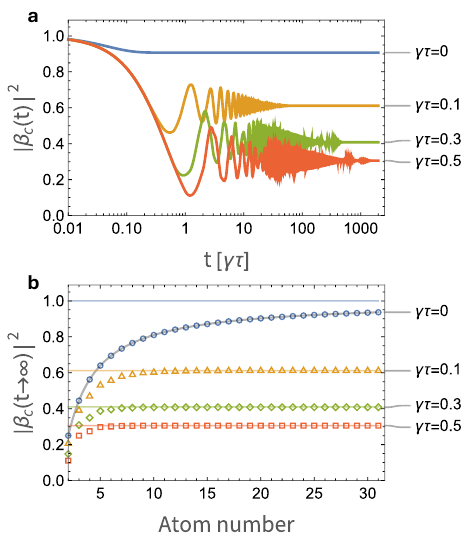}
\caption{\textbf{a} Decay dynamics of the central (initially excited) atom for various delayed interaction times $\tau$ and $N=21$ atoms. Each curve shows how the excitation probability evolves to a steady state.
\textbf{b} Steady-state excitation probability of the central atom as a function of the number of atoms in the system for different delay times. Curves correspond to distinct delays; the gray curve is the analytic Markovian solution and the horizontal dashed line marks its infinite-atom limit.}
    \label{fig2:central_atom_dynamics}
\end{figure}

{\it Central-atom dynamics.---} We analyze the dynamics of the initially excited atom located at the center of the atomic array under the Bragg condition. Figure \ref{fig2:central_atom_dynamics}\textbf{a} shows the excitation probability of the central atom ($|\beta_c(t)|^2$), which exhibits temporal oscillations before reaching a steady state. In the Markovian regime (zero delay, $\gamma\tau=0$), the steady-state excitation probability increases with the number of atoms, reflecting the collective nature of the interaction, where the excitation partially spreads across the entire array. Specifically, the central atom retains a larger fraction of the excitation as the system size increases, asymptotically reaching a probability of one for a large number of atoms in the system. The analytic expression $|\beta_c(t\rightarrow\infty)|^2=((N-1)/N)^2$ is superimposed on the Markovian case $\gamma \tau=0$ in Fig. \ref{fig2:central_atom_dynamics} \textbf{b}) (see Supplemental Material at \cite{supp} for a derivation).

In contrast, when finite delay times $\tau$ are included, the steady-state excitation probability of the central atom decreases with increasing delay and rapidly saturates as the number of atoms increases. This saturation indicates that the delayed photon-mediated interactions effectively localize a fraction of the excitation near the initially excited atom, thereby limiting the influence of the distant atoms. Hence, the excitation dynamics become predominantly local, governed by the nearest neighbors rather than the full atomic ensemble.

This behavior highlights the fundamental difference between instantaneous and delayed interactions in waveguide QED arrays: whereas instantaneous interactions promote collective excitation sharing, delay-induced effects lead to partial excitation localization and suppression of long-range correlations.

\begin{figure}
    \centering
    \includegraphics[width=1\linewidth]{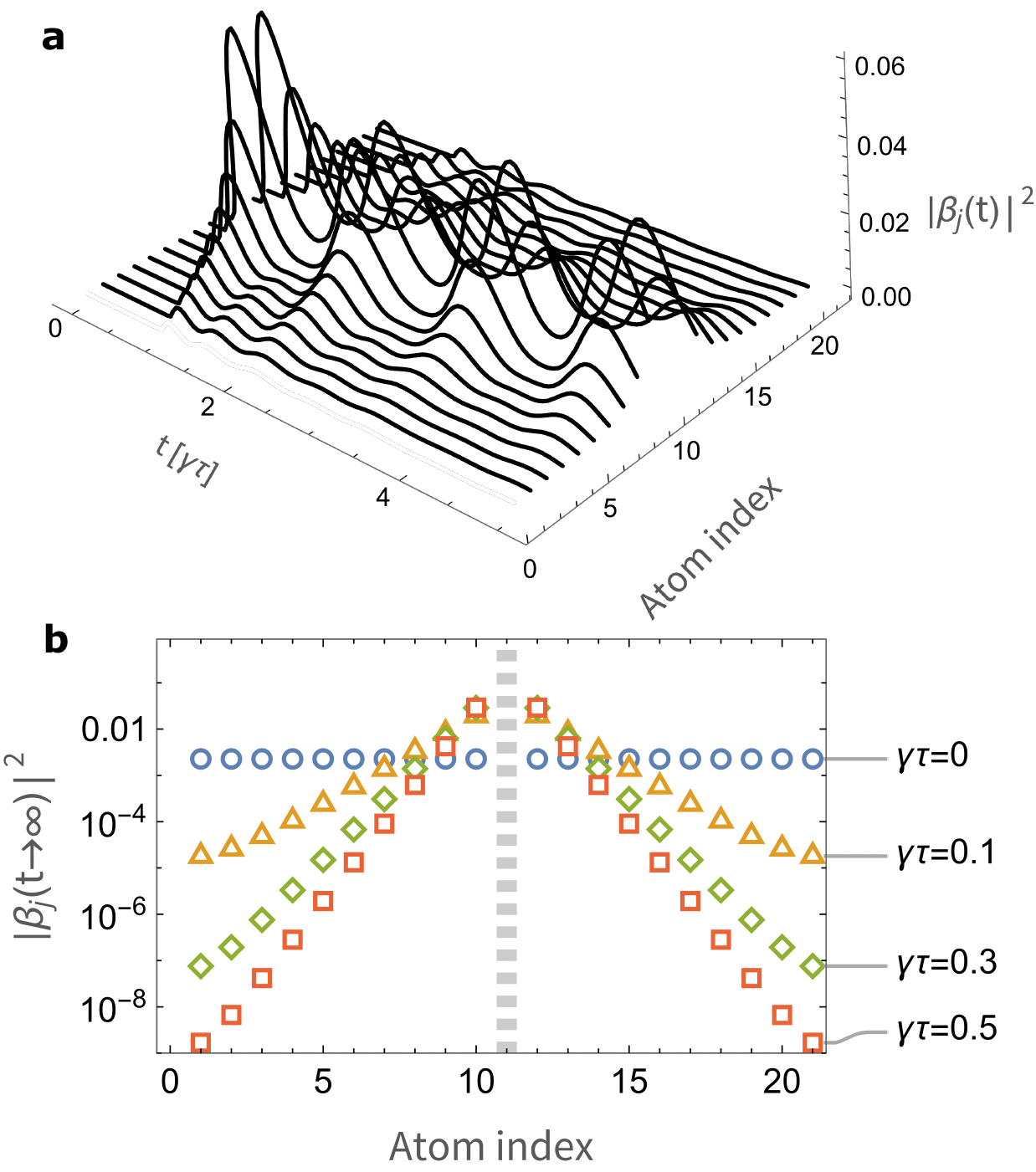}
    \caption{\textbf{a} Dynamics of the excitation probability for all neighboring atoms (excluding the central atom) as a function of time for  $\gamma \tau = 0.1$ and $N=21$ atoms. \textbf{b} Steady-state excitation probability of the neighboring atoms versus atom index for different values of $\gamma \tau$, illustrating the spatial localization of the remaining excitation around the initially excited atom.}
    \label{fig3:neighbor_atoms_dynamics}
\end{figure}

\vspace{0.1cm}
{\it Neighboring atoms dynamics.---} We now examine the excitation dynamics of all atoms in the array under the Bragg condition, excluding the initially excited central atom. Figure \ref{fig3:neighbor_atoms_dynamics} shows the dynamics of these neighboring atoms, exhibiting oscillatory behavior in their excitation probabilities as they evolve over time before settling into a steady state.

In the Markovian regime ($\gamma\tau=0$), the excitation is distributed evenly among all atoms, resulting in equal steady-state excitation probabilities $|\beta_j|^2=\left(\frac{1}{N}\right)^2$ across the array. However, when finite delay times are introduced, the excitation dynamics change significantly. The excitation probability of the neighboring atoms is spatially localized around the central atom. Moving away from the center, the excitation probability decreases exponentially, indicating strong localization effects induced by delayed photon-mediated interactions and emphasizing the role of the nearest neighbors in defining the system dynamics.

{\it Correlation length.---} To quantify the localization of atom-atom correlations in the non-Markovian regime, we analyze the spin-spin correlation function defined as \begin{equation} g(j) = \langle \sigma^{(z)}_{0} \sigma^{(z)}_{j} \rangle - \langle \sigma^{(z)}_{0} \rangle \langle \sigma^{(z)}_{j} \rangle, \end{equation} where $\sigma^{(z)}_j$ is the Pauli operator for the atom at site $j$ and the index $0$ refers to the initially excited central atom in the array.

\begin{figure}
    \centering
    \includegraphics[width=1\linewidth]{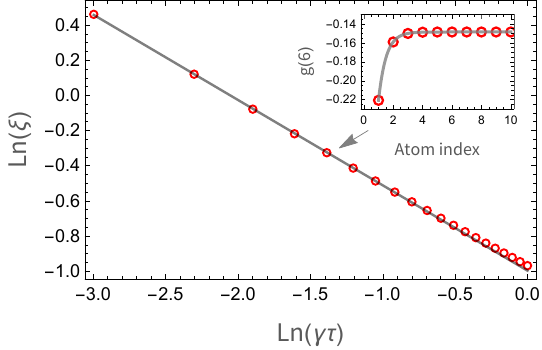}
    \caption{Correlation length $\xi$ as a function of the delayed interaction parameter $\gamma \tau$ shown in a log-log plot. The gray curve illustrates a linear fit, showing that $\xi \sim (\gamma \tau)^{-0.48}$, which suggests that the correlation length scales with the delay according to a power law. The inset shows the exponential fit to the spin-spin correlation function for the point indicated by an arrow as an example. By employing a sequence of 21 atoms, the simulation ensures that the solution converges and that boundary effects are eliminated.}
    \label{fig4:correlation_lenght}
\end{figure}

We observe that the correlation function decays exponentially with the distance from the central atom, which can be expressed as
\begin{equation} 
g(j) \propto e^{-\frac{|j|}{\xi}},
\end{equation}
where $\xi$ is the correlation length (in units of the lattice site). We extract the correlation length $\xi$ for different values of delay $\tau$ by fitting the numerical results of $g(j)$ to an exponential decay function. Figure \ref{fig4:correlation_lenght} shows that the correlation length decreases as the delay increases, indicating a stronger localization of the excitation near the initially excited atom for larger delays.

Moreover, the correlation length exhibits a power-law scaling with the delay time, given by $\xi \propto \tau^{-1/2}$. This scaling highlights the fundamental role of retardation effects in governing the correlation transport and confinement in wQED. This relationship can be derived from the equations of motion, Eq. (\ref{eq:mainresult}), as a particular manifestation of the Bragg condition. Specifically, by substituting an evanescently localized ansatz into the equations of motion, one finds that the localization length scales as $\xi \propto (\gamma\tau)^{-1/2}$ (see Supplemental Material~\cite{supp} for the full derivation). This behavior reveals that the localization effects stem from the interplay between the interaction delay time $\tau$ and the temporal coherence length of the spontaneously emitted photons, given by $\gamma^{-1}$

We notice that the atom-atom coherence function takes the form $g_{i,j}^{(+-)}=\langle \sigma^{-}_{i} \sigma^{+}_{j}\rangle- \langle \sigma^{-}_{i}\rangle\langle \sigma^{+}_{j}\rangle=\beta_i^*\beta_j$, and therefore exhibits the same power-law scaling as the spin-spin correlation function. Likewise, the two-spin entanglement, quantified by the concurrence \cite{PhysRevLett.80.2245}, is given by $C_{i,j}=|\beta_i||\beta_j|$, and thus exhibits the same scaling behavior. These similarities are expected in the context of a single excitation, where correlations arise directly from the spatial density profile of a single particle.

\begin{figure}
    \centering
    \includegraphics[width=1\linewidth]{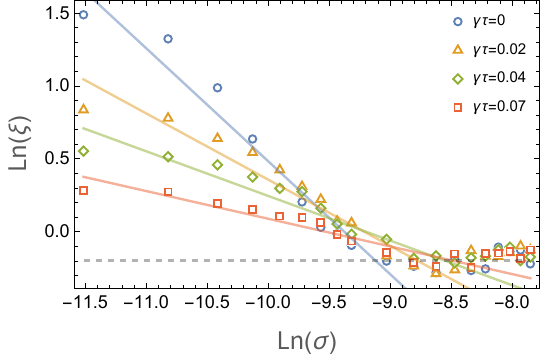}
    \caption{Correlation length $\xi$ as a function of the disorder parameter $\sigma$ shown in a log-log plot. The curves illustrate linear fits, serving as a visual aid, given by $\xi_\sigma \sim \sigma^{-0.75}$, $\sigma^{-0.5}$, $\sigma^{-0.3}$, and $\sigma^{-0.2}$ with increasing $\tau$. The gray dashed line indicates the approximatzed localization threshold of disorder.}
    \label{fig5:correlation_lenght_disorder}
\end{figure}

{\it Effects of positional disorder.---} We investigate the impact of spatial disorder in the atomic lattice on excitation localization by analyzing the correlation length $\xi$ as a function of the disorder strength $\sigma$, which quantifies the standard distribution of Gaussian random deviations from the ideal ($n\lambda$) interatomic spacing. Arbitrarily close to the Bragg condition, the atomic excitation always decays at infinitely  long times; however, this decay is slow enough to allow us to compute the correlation lengths at long times as a quasi-stationary process. The numerical simulations, shown in Fig. \ref{fig5:correlation_lenght_disorder}, reveal that even weak disorder significantly reduces the correlation length. This behavior resembles Anderson localization arising from disorder in the off-diagonal (coupling) terms of the effective Hamiltonian governing the system \cite{Anderson_1958}. 

For small disorders, the correlation length decreased approximately following a power-law scaling, similar to that observed with increasing delay. However, unlike the delay-induced case, $\xi$ saturates at a finite minimum value slightly below the size of a single lattice site. This saturation defines a localization limit imposed by disorder, below which further increases in disorder do not reduce the spatial extent of the correlation localization. Furthermore, the interplay between the delay and disorder modifies the scaling behavior of the correlation length. 

The findings indicate that delay-induced non-Markovian interactions and spatial disorder have independent and combined effects on correlation localization within wQED arrays. Retardation effects result in a greater confinement of correlations than spatial randomness; however, even weak disorder establishes a lower bound on the correlation length. 

{\it Discussion and Outlook.---} 
The steady-state localization of excitation presented here critically depends on the Bragg condition combined with a perfect emitter-waveguide coupling. Deviations from the Bragg condition or imperfect coupling introduce additional decay channels into the radiated modes, preventing a true steady state and leading to eventual excitation loss. Nonetheless, when these decay processes occur sufficiently slowly, the transient dynamics still exhibit the characteristic localization effects described here, albeit only during out-of-equilibrium evolution. Quantifying and comparing such transient quasi-localizations across varying system parameters remains challenging owing to the lack of a straightforward metric for nonsteady-state behaviors.

The localization of atomic excitation with delayed interactions implies a nonzero field probability amplitude confined within partially excited atoms. This resembles BIC states, which have been previously studied in simpler two-atom and giant atom systems \cite{pichler2016,ciccarello_2026,solano_2020_dist,Sinha_2020,Giuseppe_2025,carmele_2020,Alvarez_2024,Calajo_2019,guo_2020,zueco_2021,xinyou_2024}. However, photonic bound states in extended atomic arrays remain unexplored. Studying these phenomena could deepen our understanding of the atomic-photonic interplay in non-Markovian wQED and enable new control strategies for light-matter interactions.

The observed localization phenomena bear a conceptual resemblance to statistical mechanics models. In the Markovian limit, the system dynamics evoke analogies to the Kuramoto model of coupled oscillators \cite{KuramotoModel}, whereas the non-Markovian regime with delayed interactions parallels the features of the Kac-Ising model, which is characterized by finite-range interactions \cite{10.1063/1.1664976}. However, extending these analogies rigorously to our  wQED system requires a formalism beyond the single-excitation manifold, as the nonlinear atomic response and many-body effects complicate the analysis. This highlights an intriguing and underexplored avenue: the thermodynamics of collective systems with delayed interactions. Addressing this direction necessitates advanced numerical tools and theoretical frameworks capable of capturing non-Markovian many-body dynamics.

Because the Markovian limit is an idealization that is never truly realized in practice, realistic wQED systems inevitably exhibit some degree of retardation and thus non-Markovianity. Even within the single-excitation regime, our results demonstrate that delay-induced interactions inherently act as isolators by partially localizing the excitation and correlations near the initially excited atom. This intrinsic localization may have practical implications for controlling excitation and correlation transport and for designing quantum devices that exploit non-Markovian effects.

{\it Conclusion.---} 
Our research demonstrates that delay-induced non-Markovian interactions in atomic arrays, when coupled to a waveguide under Bragg conditions, result in the localization of atomic correlations within the single-excitation regime. This phenomenon is quantitatively characterized by the spin-spin correlation function, which exhibits exponential decay with respect to the atomic site distance. This decay facilitates the determination of a correlation length that scales with the square root of the inverse of the photon travel delay time. Our results indicate that retardation effects fundamentally modify the excitation transport by suppressing long-range coherence and effectively confining the correlations to finite spatial regions.

{\it Acknowledgments.---} P.S. thanks K. Sinha for helpful discussions. P.B.-B. acknowledges support from the PAPIIT-DGAPA Grant No. IG101324. P.S., N.V., and F.Q. acknowledge support from the National Agency for Research and Development (ANID) through Project FONDECYT Grants No. 1240204, ANID-Subdirecci\'on de Capital Humano/Doctorado Nacional/2022-21221251 and ANID-Subdirecci\'on de Capital Humano/Doctorado Nacional/2024-21242548, respectively.

\bibliography{ref}

@article{chang2018,                             
  author  = {Chang, D. E. and Douglas, J. S. and Gonz{\'a}lez-Tudela, A. and Hung, C.-L. and Kimble, H. J.},
  title   = {Colloquium: Quantum matter built from nanoscopic lattices of atoms and photons},
  journal = {Rev. Mod. Phys.},
  volume  = {90},
  pages   = {031002},
  year    = {2018},
  doi     = {10.1103/RevModPhys.90.031002}
}

@article{sheremet2023,          
  author  = {Sheremet, Alexandra S. and Petrov, Mihail I. and Iorsh, Ivan V. and Poshakinskiy, Alexander V. and Poddubny, Alexander N.},
  title   = {Waveguide quantum electrodynamics: collective radiance and photon--photon correlations},
  journal = {Rev. Mod. Phys.},
  volume  = {95},
  pages   = {015002},
  year    = {2023},
  doi     = {10.1103/RevModPhys.95.015002}
}

@article{roy2017,                              
  author  = {Roy, Dibyendu and Wilson, C. M. and Firstenberg, Ofer},
  title   = {Colloquium: Strongly interacting photons in one-dimensional continuum},
  journal = {Rev. Mod. Phys.},
  volume  = {89},
  pages   = {021001},
  year    = {2017},
  doi     = {10.1103/RevModPhys.89.021001}
}

@article{solano2017optical,
author = {Solano, P. and Grover, J. A. and Hoffman, J. E. and Ravets, S.
and Fatemi, F. K. and Orozco, L. A. and Rolston, S. L.},
title = {Optical nanofibers: A new platform for quantum optics},
journal = {Adv. At. Mol. Opt. Phys.},
volume = {66}, pages = {439--505}, year = {2017},
doi = {10.1016/bs.aamop.2017.02.003}}

@article{NatureCorzo2019,
  title = {Waveguide-coupled single collective excitation of atomic arrays},
  author = {Corzo, Neil V. and Raskop, Jérémy and Chandra, Aveek and Sheremet, Alexandra S and Gouraud, Baptiste and Laurat, Julien},
  journal = {Nature},
  volume = {566},
  issue = {7744},
  pages = {359-362},
  year = {2019},
  month = {Feb},
  doi = {10.1038/s41586-019-0902-3},
  url = {https://doi.org/10.1038/s41586-019-0902-3}
}

@article{Paulisch_2016,
doi = {10.1088/1367-2630/18/4/043041},
url = {https://doi.org/10.1088/1367-2630/18/4/043041},
year = {2016},
month = {apr},
publisher = {IOP Publishing},
volume = {18},
number = {4},
pages = {043041},
author = {Paulisch, V and Kimble, H J and González-Tudela, A},
title = {Universal quantum computation in waveguide QED using decoherence free subspaces},
journal = {New Journal of Physics}
}

@article{kimble_2008, author={Kimble, H. J.}, title={The quantum
internet}, journal={Nature}, volume={453}, pages={1023-1030},
year={2008}, doi={10.1038/nature07127}}

@article{milonni1995photodetection,
  title={Photodetection and causality in quantum optics},
  author={Milonni, PW and James, DFV and Fearn, Heidi},
  journal={Physical Review A},
  volume={52},
  number={2},
  pages={1525},
  year={1995},
  publisher={APS}
}

@article{Sinha_2020,
  title = {Non-Markovian Collective Emission from Macroscopically Separated Emitters},
  author = {Sinha, Kanupriya and Meystre, Pierre and Goldschmidt, Elizabeth A. and Fatemi, Fredrik K. and Rolston, S. L. and Solano, Pablo},
  journal = {Phys. Rev. Lett.},
  volume = {124},
  issue = {4},
  pages = {043603},
  numpages = {7},
  year = {2020},
  month = {Jan},
  publisher = {American Physical Society},
  doi = {10.1103/PhysRevLett.124.043603},
  url = {https://link.aps.org/doi/10.1103/PhysRevLett.124.043603}
}

@article{Giuseppe_2025,
  title = {Non-Markovian dynamics of generation of bound states in the continuum via single-photon scattering},
  author = {Magnifico, Giuseppe and Maffei, Maria and Pomarico, Domenico and Das, Debmalya and Facchi, Paolo and Pascazio, Saverio and Pepe, Francesco V.},
  journal = {Phys. Rev. Res.},
  volume = {7},
  issue = {3},
  pages = {033249},
  numpages = {9},
  year = {2025},
  month = {Sep},
  publisher = {American Physical Society},
  doi = {10.1103/gtf6-zb57},
  url = {https://link.aps.org/doi/10.1103/gtf6-zb57}
}

@article{Alvarez_2024,
  title = {Delay-induced spontaneous dark-state generation from two distant excited atoms},
  author = {Alvarez-Giron, W. and Solano, P. and Sinha, K. and Barberis-Blostein, P.},
  journal = {Phys. Rev. Res.},
  volume = {6},
  issue = {2},
  pages = {023213},
  numpages = {13},
  year = {2024},
  month = {May},
  publisher = {American Physical Society},
  doi = {10.1103/PhysRevResearch.6.023213},
  url = {https://link.aps.org/doi/10.1103/PhysRevResearch.6.023213}
}

@article{Calajo_2019,
  title = {Exciting a Bound State in the Continuum through Multiphoton Scattering Plus Delayed Quantum Feedback},
  author = {Calaj\'o, Giuseppe and Fang, Yao-Lung L. and Baranger, Harold U. and Ciccarello, Francesco},
  journal = {Phys. Rev. Lett.},
  volume = {122},
  issue = {7},
  pages = {073601},
  numpages = {7},
  year = {2019},
  month = {Feb},
  publisher = {American Physical Society},
  doi = {10.1103/PhysRevLett.122.073601},
  url = {https://link.aps.org/doi/10.1103/PhysRevLett.122.073601}
}

@article{Tecer_2026,
  title = {Flat-band-mediated photon-photon interactions in two-dimensional waveguide QED networks},
  author = {Te\ifmmode \check{c}\else \v{c}\fi{}er, Matija and Calaj\'o, Giuseppe and Di Liberto, Marco},
  journal = {Phys. Rev. A},
  volume = {113},
  issue = {1},
  pages = {013701},
  numpages = {12},
  year = {2026},
  month = {Jan},
  publisher = {American Physical Society},
  doi = {10.1103/dt14-h2c2},
  url = {https://link.aps.org/doi/10.1103/dt14-h2c2}
}

@article{Ciccarello_2024Wqed,
  author  = {Ciccarello, Francesco and Lodahl, Peter and Schneble, Dominik},
  title   = {Waveguide Quantum Electrodynamics},
  journal = {Optics \& Photonics News},
  volume  = {35},
  number  = {1},
  pages   = {34--41},
  year    = {2024},
  doi     = {10.1364/OPN.35.1.000034},
  url     = {https://doi.org/10.1364/OPN.35.1.000034}
}

@article{Lindblad1976,
  author  = {Lindblad, G{\"o}ran},
  title   = {On the generators of quantum dynamical semigroups},
  journal = {Communications in Mathematical Physics},
  volume  = {48},
  number  = {2},
  pages   = {119--130},
  year    = {1976},
  doi     = {10.1007/BF01608499}
}

@article{Lehmberg1970,
  author  = {Lehmberg, R. H.},
  title   = {Radiation from an $N$-Atom System. I. General Formalism},
  journal = {Phys. Rev. A},
  volume  = {2},
  number  = {3},
  pages   = {883--888},
  year    = {1970},
  doi     = {10.1103/PhysRevA.2.883}
}

@article{Carmichael1993,
  author  = {Carmichael, H. J.},
  title   = {Quantum trajectory theory for cascaded open systems},
  journal = {Phys. Rev. Lett.},
  volume  = {70},
  number  = {15},
  pages   = {2273--2276},
  year    = {1993},
  doi     = {10.1103/PhysRevLett.70.2273}
}

@article{Asenjo_2017,
  title = {Exponential Improvement in Photon Storage Fidelities Using Subradiance and ``Selective Radiance'' in Atomic Arrays},
  author = {Asenjo-Garcia, A. and Moreno-Cardoner, M. and Albrecht, A. and Kimble, H. J. and Chang, D. E.},
  journal = {Phys. Rev. X},
  volume = {7},
  issue = {3},
  pages = {031024},
  year = {2017},
  month = {Aug},
  publisher = {American Physical Society},
  doi = {10.1103/PhysRevX.7.031024},
  url = {https://link.aps.org/doi/10.1103/PhysRevX.7.031024}
}

@article{schilke2011,
  title = {Photonic Band Gaps in One-Dimensionally Ordered Cold Atomic Vapors},
  author = {Schilke, Alexander and Zimmermann, Claus and Courteille, Philippe W. and Guerin, William},
  journal = {Phys. Rev. Lett.},
  volume = {106},
  issue = {22},
  pages = {223903},
  year = {2011},
  month = {Jun},
  publisher = {American Physical Society},
  doi = {10.1103/PhysRevLett.106.223903},
  url = {https://link.aps.org/doi/10.1103/PhysRevLett.106.223903}
}

@article{Kimble_2015,
  title = {Superradiance for Atoms Trapped along a Photonic Crystal Waveguide},
  author = {Goban, A. and Hung, C.-L. and Hood, J. D. and Yu, S.-P. and Muniz, J. A. and Painter, O. and Kimble, H. J.},
  journal = {Phys. Rev. Lett.},
  volume = {115},
  issue = {6},
  pages = {063601},
  year = {2015},
  month = {Aug},
  publisher = {American Physical Society},
  doi = {10.1103/PhysRevLett.115.063601},
  url = {https://link.aps.org/doi/10.1103/PhysRevLett.115.063601}
}

@article{Olmos_2021,
  title = {Bragg condition for scattering into a guided optical mode},
  author = {Olmos, B. and Liedl, C. and Lesanovsky, I. and Schneeweiss, P.},
  journal = {Phys. Rev. A},
  volume = {104},
  issue = {4},
  pages = {043517},
  year = {2021},
  month = {Oct},
  publisher = {American Physical Society},
  doi = {10.1103/PhysRevA.104.043517},
  url = {https://link.aps.org/doi/10.1103/PhysRevA.104.043517}
}

@article{Nieddu2016Jjop,
doi = {10.1088/2040-8978/18/5/053001},
url = {https://dx.doi.org/10.1088/2040-8978/18/5/053001},
year = {2016},
month = {mar},
publisher = {IOP Publishing},
volume = {18},
number = {5},
pages = {053001},
author = {Nieddu, Thomas and Gokhroo, Vandna and Nic Chormaic, Síle },
title = {Optical nanofibres and neutral atoms},
journal = {Journal of Optics},
}

@article{cardenas2023many,
  title = {Many-Body Superradiance and Dynamical Mirror Symmetry Breaking in Waveguide QED},
  author = {Cardenas-Lopez, Silvia and Masson, Stuart J. and Zager, Zoe and Asenjo-Garcia, Ana},
  journal = {Phys. Rev. Lett.},
  volume = {131},
  issue = {3},
  pages = {033605},
  year = {2023},
  month = {Jul},
  publisher = {American Physical Society},
  doi = {10.1103/PhysRevLett.131.033605},
  url = {https://link.aps.org/doi/10.1103/PhysRevLett.131.033605}
}

@article{Solano2017_inf,
  author  = {Solano, Pablo and Barberis-Blostein, Pablo and Fatemi, Fredrik K. and Orozco, Luis A. and Rolston, Steven L.},
  title   = {Super-radiance reveals infinite-range dipole interactions through a nanofiber},
  journal = {Nature Communications},
  volume  = {8},
  pages   = {1857},
  year    = {2017},
  doi     = {10.1038/s41467-017-01994-3},
  url     = {https://doi.org/10.1038/s41467-017-01994-3}
}

@article{Abanin_2019,
  title = {Colloquium: Many-body localization, thermalization, and entanglement},
  author = {Abanin, Dmitry A. and Altman, Ehud and Bloch, Immanuel and Serbyn, Maksym},
  journal = {Rev. Mod. Phys.},
  volume = {91},
  issue = {2},
  pages = {021001},
  numpages = {26},
  year = {2019},
  month = {May},
  publisher = {American Physical Society},
  doi = {10.1103/RevModPhys.91.021001},
  url = {https://link.aps.org/doi/10.1103/RevModPhys.91.021001}
}

@article{Annyun_2025,
  title = {Non-Markovian spontaneous emission in a tunable cavity formed by atomic mirrors},
  author = {Das, Annyun and Solano, Pablo and Sinha, Kanu},
  journal = {Phys. Rev. A},
  volume = {112},
  issue = {4},
  pages = {043723},
  numpages = {11},
  year = {2025},
  month = {Oct},
  publisher = {American Physical Society},
  doi = {10.1103/29yv-12sq},
  url = {https://link.aps.org/doi/10.1103/29yv-12sq}
}

@article{Sinha_2025,
doi = {10.1088/1367-2630/add495},
url = {https://doi.org/10.1088/1367-2630/add495},
year = {2025},
month = {may},
publisher = {IOP Publishing},
volume = {27},
number = {5},
pages = {054101},
author = {Sinha, Kanu and Parra-Contreras, Jennifer and Das, Annyun and Solano, Pablo},
title = {Spontaneous emission in the presence of quantum mirrors},
journal = {New Journal of Physics},
}

@article{chang2012,
  title = {Cavity QED with atomic mirrors},
  author = {D E Chang and L Jiang and A V Gorshkov and H J Kimble},
  journal = {New Journal of Physics},
  volume = {14},
  number = {6},
  pages = {063003},
  year = {2012},
  month = {Jun},
  publisher = {IOP Publishing},
  doi = {10.1088/1367-2630/14/6/063003},
  url = {https://dx.doi.org/10.1088/1367-2630/14/6/063003}
}

@misc{kanu_2026,
      title={Non-Markovian delay-assisted sensing with waveguide-coupled quantum emitters}, 
      author={Prajit Dhara and Isack Padilla and Saikat Guha and Annyun Das and Kanu Sinha},
      year={2026},
      eprint={2605.05434},
      archivePrefix={arXiv}
}

@article{ciraczoller1997,                       
author  = {Cirac, J. I. and Zoller, P. and Kimble, H. J. and Mabuchi, H.},
  title   = {Quantum State Transfer and Entanglement Distribution among Distant Nodes in a Quantum Network},
  journal = {Phys. Rev. Lett.},
  volume  = {78},
  pages   = {3221--3224},
  year    = {1997},
  doi     = {10.1103/PhysRevLett.78.3221}
}

@article{lechner2023,                          
  author  = {Lechner, Daniel and Pennetta, Riccardo and Blaha, Martin and Schneeweiss, Philipp and Rauschenbeutel, Arno and Volz, J{\"u}rgen},
  title   = {Light-Matter Interaction at the Transition Between Cavity and Waveguide QED},
  journal = {Phys. Rev. Lett.},
  volume  = {131},
  pages   = {103603},
  year    = {2023},
  doi     = {10.1103/PhysRevLett.131.103603}
}

@article{Dominik_2024,
  title = {Exact solution for the collective non-Markovian decay of two fully excited quantum emitters},
  author = {Lanuza, Alfonso and Schneble, Dominik},
  journal = {Phys. Rev. Res.},
  volume = {6},
  issue = {3},
  pages = {033196},
  numpages = {16},
  year = {2024},
  month = {Aug},
  publisher = {American Physical Society},
  doi = {10.1103/PhysRevResearch.6.033196},
  url = {https://link.aps.org/doi/10.1103/PhysRevResearch.6.033196}
}

@article{Cirac_2025,
  title = {Effects of Retardation on Many-Body Superradiance in Chiral Waveguide QED},
  author = {Windt, Bennet and Bello, Miguel and Malz, Daniel and Cirac, J. Ignacio},
  journal = {Phys. Rev. Lett.},
  volume = {134},
  issue = {17},
  pages = {173601},
  numpages = {8},
  year = {2025},
  month = {Apr},
  publisher = {American Physical Society},
  doi = {10.1103/PhysRevLett.134.173601},
  url = {https://link.aps.org/doi/10.1103/PhysRevLett.134.173601}
}

@article{douglas2015,                           
  author  = {Douglas, J. S. and Habibian, H. and Hung, C.-L. and Gorshkov, A. V. and Kimble, H. J. and Chang, D. E.},
  title   = {Quantum many-body models with cold atoms coupled to photonic crystals},
  journal = {Nat. Photonics},
  volume  = {9},
  pages   = {326--331},
  year    = {2015},
  doi     = {10.1038/nphoton.2015.57}
}

@article{ferreira2021,                         
author  = {Ferreira, V{\'i}ctor S. and Banker, Jash and Sipahigil, Alp and Matheny, Matthew H. and Keller, Andrew J. and Kim, Eunjong and Mirhosseini, Mohammad and Painter, Oskar},
  title   = {Collapse and Revival of an Artificial Atom Coupled to a Structured Photonic Reservoir},
  journal = {Phys. Rev. X},
  volume  = {11},
  pages   = {041043},
  year    = {2021},
  doi     = {10.1103/PhysRevX.11.041043}
}

@misc{hughes_2026,
      title={Resonance fluorescence of an artificial atom with a time-delayed coherent feedback}, 
      author={Ching-Yeh Chen and Gavin Crowder and Zheng-Qi Niu and Ping Yi Wen and Yen-Hsiang Lin and Jeng-Chung Chen and Zhi-Rong Lin and Franco Nori and Stephen Hughes and Io-Chun Hoi},
      year={2026},
      eprint={2603.28004},
      archivePrefix={arXiv}
}

@misc{mohammad_2025,
      title={Long-distance cascaded fluorescence of cold Cesium atoms coupled to an optical nanofiber}, 
      author={Mohammad Sadeghi and Wayne Crump and Scott Parkins and Maarten Hoogerland},
      year={2025},
      eprint={2412.01099},
      archivePrefix={arXiv} 
}

@article{pichler2016,      
  author  = {Pichler, Hannes and Zoller, Peter},
  title   = {Photonic Circuits with Time Delays and Quantum Feedback},
  journal = {Phys. Rev. Lett.},
  volume  = {116},
  pages   = {093601},
  year    = {2016},
  doi     = {10.1103/PhysRevLett.116.093601}
}

@article{pichler2017,  
  author  = {Pichler, Hannes and Choi, Soonwon and Zoller, Peter and Lukin, Mikhail D.},
  title   = {Universal photonic quantum computation via time-delayed feedback},
  journal = {Proc. Natl. Acad. Sci. U.S.A.},
  volume  = {114},
  number  = {43},
  pages   = {11362--11367},
  year    = {2017},
  doi     = {10.1073/pnas.1711003114}
}

@article{Ramos_2016,
  title = {Non-Markovian dynamics in chiral quantum networks with spins and photons},
  author = {Ramos, Tom\'as and Vermersch, Beno\^{\i}t and Hauke, Philipp and Pichler, Hannes and Zoller, Peter},
  journal = {Phys. Rev. A},
  volume = {93},
  issue = {6},
  pages = {062104},
  numpages = {23},
  year = {2016},
  month = {Jun},
  publisher = {American Physical Society},
  doi = {10.1103/PhysRevA.93.062104},
  url = {https://link.aps.org/doi/10.1103/PhysRevA.93.062104}
}

@article{zheng_2013qc,                      
  author  = {Zheng, Huaixiu and Gauthier, Daniel J. and Baranger, Harold U.},
  title   = {Waveguide-QED-Based Photonic Quantum Computation},
  journal = {Phys. Rev. Lett.},
  volume  = {111},
  pages   = {090502},
  year    = {2013},
  doi     = {10.1103/PhysRevLett.111.090502}
}

@article{guo_2020,                               
  author  = {Guo, Lingzhen and Kockum, Anton Frisk and Marquardt, Florian and     Johansson, G{\"o}ran},
  title   = {Oscillating bound states for a giant atom},
  journal = {Phys. Rev. Res.},
  volume  = {2},
  pages   = {043014},
  year    = {2020},
  doi     = {10.1103/PhysRevResearch.2.043014}
}

@article{zheng2013,                
  title = {Persistent Quantum Beats and Long-Distance Entanglement from Waveguide-Mediated Interactions},
  author = {Zheng, Huaixiu and Baranger, Harold U.},
  journal = {Phys. Rev. Lett.},
  volume = {110},
  issue = {11},
  pages = {113601},
  numpages = {5},
  year = {2013},
  month = {Mar},
  publisher = {American Physical Society},
  doi = {10.1103/PhysRevLett.110.113601},
  url = {https://link.aps.org/doi/10.1103/PhysRevLett.110.113601}
}

@article{ciccarello_2026,
  title = {Multimode-cavity picture of non-Markovian waveguide QED},
  author = {Cilluffo, Dario and Ferialdi, Luca and Palma, G. Massimo and Calaj\`o, Giuseppe and Ciccarello, Francesco},
  journal = {Phys. Rev. Res.},
  volume = {8},
  issue = {2},
  pages = {023172},
  numpages = {14},
  year = {2026},
  month = {May},
  publisher = {American Physical Society},
  doi = {10.1103/hc4r-2st8},
  url = {https://link.aps.org/doi/10.1103/hc4r-2st8}
}

@misc{maffei_2025,
      title={Superradiant decay in non-Markovian Waveguide Quantum Electrodynamics}, 
      author={Rosa Lucia Capurso and Giuseppe Calajó and Simone Montangero and Saverio Pascazio and Francesco V. Pepe and Maria Maffei and Giuseppe Magnifico and Paolo Facchi},
      year={2025},
      eprint={2511.22332},
      archivePrefix={arXiv}
}

@article{Wallraff_2022,
  title = {Controlling Atom-Photon Bound States in an Array of Josephson-Junction Resonators},
  author = {Scigliuzzo, Marco and Calaj\`o, Giuseppe and Ciccarello, Francesco and Perez Lozano, Daniel and Bengtsson, Andreas and Scarlino, Pasquale and Wallraff, Andreas and Chang, Darrick and Delsing, Per and Gasparinetti, Simone},
  journal = {Phys. Rev. X},
  volume = {12},
  issue = {3},
  pages = {031036},
  numpages = {22},
  year = {2022},
  month = {Sep},
  publisher = {American Physical Society},
  doi = {10.1103/PhysRevX.12.031036},
  url = {https://link.aps.org/doi/10.1103/PhysRevX.12.031036}
}

@article{xinyou_2024,
  title = {Non-Markovian collective emission of giant emitters in the Zeno regime},
  author = {Qiu, Qing-Yang and L\"u, Xin-You},
  journal = {Phys. Rev. Res.},
  volume = {6},
  issue = {3},
  pages = {033243},
  numpages = {13},
  year = {2024},
  month = {Sep},
  publisher = {American Physical Society},
  doi = {10.1103/PhysRevResearch.6.033243},
  url = {https://link.aps.org/doi/10.1103/PhysRevResearch.6.033243}
}

@article{dissimilar_2023,
  title = {Dissimilar collective decay and directional emission from two quantum emitters},
  author = {Solano, P. and Barberis-Blostein, P. and Sinha, K.},
  journal = {Phys. Rev. A},
  volume = {107},
  issue = {2},
  pages = {023723},
  numpages = {14},
  year = {2023},
  month = {Feb},
  publisher = {American Physical Society},
  doi = {10.1103/PhysRevA.107.023723},
  url = {https://link.aps.org/doi/10.1103/PhysRevA.107.023723}
}

@article{zueco_2021,                 
  author  = {Gonz{\'a}lez-Guti{\'e}rrez, C. A. and Rom{\'a}n-Roche, J. and Zueco, D.},
  title   = {Distant emitters in ultrastrong waveguide QED: Ground-state properties and non-Markovian dynamics},
  journal = {Phys. Rev. A},
  volume  = {104},
  pages   = {053701},
  year    = {2021},
  doi     = {10.1103/PhysRevA.104.053701}
}

@article{carmele_2020,
  title = {Pronounced non-Markovian features in multiply excited, multiple emitter waveguide QED: Retardation induced anomalous population trapping},
  author = {Carmele, Alexander and Nemet, Nikolett and Canela, Victor and Parkins, Scott},
  journal = {Phys. Rev. Res.},
  volume = {2},
  issue = {1},
  pages = {013238},
  numpages = {9},
  year = {2020},
  month = {Mar},
  publisher = {American Physical Society},
  doi = {10.1103/PhysRevResearch.2.013238},
  url = {https://link.aps.org/doi/10.1103/PhysRevResearch.2.013238}
}

@article{solano_2020_dist,
  title = {Collective radiation from distant emitters},
  author = {Sinha, Kanupriya and Gonz\'alez-Tudela, Alejandro and Lu, Yong and Solano, Pablo},
  journal = {Phys. Rev. A},
  volume = {102},
  issue = {4},
  pages = {043718},
  numpages = {12},
  year = {2020},
  month = {Oct},
  publisher = {American Physical Society},
  doi = {10.1103/PhysRevA.102.043718},
  url = {https://link.aps.org/doi/10.1103/PhysRevA.102.043718}
}

@article{arno_2010,
  title = {Optical Interface Created by Laser-Cooled Atoms Trapped in the Evanescent Field Surrounding an Optical Nanofiber},
  author = {Vetsch, E. and Reitz, D. and Sagu\'e, G. and Schmidt, R. and Dawkins, S. T. and Rauschenbeutel, A.},
  journal = {Phys. Rev. Lett.},
  volume = {104},
  issue = {20},
  pages = {203603},
  numpages = {4},
  year = {2010},
  month = {May},
  publisher = {American Physical Society},
  doi = {10.1103/PhysRevLett.104.203603},
  url = {https://link.aps.org/doi/10.1103/PhysRevLett.104.203603}
}

@article{KuramotoModel,
  title = {The Kuramoto model: A simple paradigm for synchronization phenomena},
  author = {Acebr\'on, Juan A. and Bonilla, L. L. and P\'erez Vicente, Conrad J. and Ritort, F\'elix and Spigler, Renato},
  journal = {Rev. Mod. Phys.},
  volume = {77},
  issue = {1},
  pages = {137--185},
  numpages = {0},
  year = {2005},
  month = {Apr},
  publisher = {American Physical Society},
  doi = {10.1103/RevModPhys.77.137},
  url = {https://link.aps.org/doi/10.1103/RevModPhys.77.137}
}

@article{PhysRevA.77.043833,
  title = {Cooperative spontaneous emission as a many-body eigenvalue problem},
  author = {Svidzinsky, Anatoly and Chang, Jun-Tao},
  journal = {Phys. Rev. A},
  volume = {77},
  issue = {4},
  pages = {043833},
  numpages = {4},
  year = {2008},
  month = {Apr},
  publisher = {American Physical Society},
  doi = {10.1103/PhysRevA.77.043833},
  url = {https://link.aps.org/doi/10.1103/PhysRevA.77.043833}
}

@article{PhysRevA.91.042116,
  title = {Quantum optics of chiral spin networks},
  author = {Pichler, Hannes and Ramos, Tom\'as and Daley, Andrew J. and Zoller, Peter},
  journal = {Phys. Rev. A},
  volume = {91},
  issue = {4},
  pages = {042116},
  numpages = {19},
  year = {2015},
  month = {Apr},
  publisher = {American Physical Society},
  doi = {10.1103/PhysRevA.91.042116},
  url = {https://link.aps.org/doi/10.1103/PhysRevA.91.042116}
}

@article{PhysRevLett.80.2245,
  title = {Entanglement of Formation of an Arbitrary State of Two Qubits},
  author = {Wootters, William K.},
  journal = {Phys. Rev. Lett.},
  volume = {80},
  issue = {10},
  pages = {2245--2248},
  numpages = {0},
  year = {1998},
  month = {Mar},
  publisher = {American Physical Society},
  doi = {10.1103/PhysRevLett.80.2245},
  url = {https://link.aps.org/doi/10.1103/PhysRevLett.80.2245}
}

@article{10.1063/1.1664976,
    author = {Kac, Mark and Thompson, Colin J.},
    title = {Critical Behavior of Several Lattice Models with Long‐Range Interaction},
    journal = {Journal of Mathematical Physics},
    volume = {10},
    number = {8},
    pages = {1373-1386},
    year = {1969},
    month = {08},
    issn = {0022-2488},
    doi = {10.1063/1.1664976},
    url = {https://doi.org/10.1063/1.1664976},
}

@article{Wallraff_qLink_2020,
  title = {Microwave Quantum Link between Superconducting Circuits Housed in Spatially Separated Cryogenic Systems},
  author = {Magnard, P. and Storz, S. and Kurpiers, P. and Sch\"ar, J. and Marxer, F. and L\"utolf, J. and Walter, T. and Besse, J.-C. and Gabureac, M. and Reuer, K. and Akin, A. and Royer, B. and Blais, A. and Wallraff, A.},
  journal = {Phys. Rev. Lett.},
  volume = {125},
  issue = {26},
  pages = {260502},
  numpages = {7},
  year = {2020},
  month = {Dec},
  publisher = {American Physical Society},
  doi = {10.1103/PhysRevLett.125.260502},
  url = {https://link.aps.org/doi/10.1103/PhysRevLett.125.260502}
}

@article{
Wallraff_2013,
author = {Arjan F. van Loo  and Arkady Fedorov  and Kevin Lalumière  and Barry C. Sanders  and Alexandre Blais  and Andreas Wallraff },
title = {Photon-Mediated Interactions Between Distant Artificial Atoms},
journal = {Science},
volume = {342},
number = {6165},
pages = {1494-1496},
year = {2013},
doi = {10.1126/science.1244324},
URL = {https://www.science.org/doi/abs/10.1126/science.1244324}
}

@article{Weisskopf1930,
  author  = {Weisskopf, V. and Wigner, E.},
  title   = {Berechnung der nat{\"u}rlichen Linienbreite auf Grund der Diracschen Lichttheorie},
  journal = {Zeitschrift f{\"u}r Physik},
  year    = {1930},
  volume  = {63},
  number  = {1},
  pages   = {54--73},
  issn    = {0044-3328},
  doi     = {10.1007/BF01336768},
  url     = {https://doi.org/10.1007/BF01336768}
}

@article{Anderson_1958,
  title = {Absence of Diffusion in Certain Random Lattices},
  author = {Anderson, P. W.},
  journal = {Phys. Rev.},
  volume = {109},
  issue = {5},
  pages = {1492--1505},
  numpages = {0},
  year = {1958},
  month = {Mar},
  publisher = {American Physical Society},
  doi = {10.1103/PhysRev.109.1492},
  url = {https://link.aps.org/doi/10.1103/PhysRev.109.1492}
}

@misc{supp,
  note = "See Supplemental Material at
    URL-will-be-inserted-by-publisher for detailed calculations."
}

\clearpage

\setcounter{equation}{0}
\setcounter{figure}{0}
\setcounter{table}{0}
\setcounter{page}{1}
\setcounter{section}{0}

\renewcommand{\theequation}{S\arabic{equation}}
\renewcommand{\thefigure}{S\arabic{figure}}
\renewcommand{\thetable}{S\arabic{table}}
\renewcommand{\thesection}{S\arabic{section}}

\section{Supplemental Material for: Delayed-Induced Excitation Localization in Atomic Arrays Coupled to a Waveguide}

\maketitle

\subsection{Derivation of Eq. (3) in main text}
\label{sec:Appendix1}

Upon replacing into the Schrödinger equation, we get two differential equations for the field and atomic coefficients,

{\small
\begin{align}
    &\dot{\beta}_j(t)=-\sum_{\lambda=L,R}\int d\omega \sqrt{\frac{\gamma_\lambda}{2\pi}}e^{-i\left[\omega x_j /v_\lambda-(\omega-\omega_0) t\right]}\alpha_\lambda (\omega,t)\label{eq:diffbeta}\\
    &\dot{\alpha}_\lambda(t,\omega)=\sum_j \sqrt{\frac{\gamma_\lambda}{2\pi}} \beta_j(t)e^{i\left[\omega x_j /v_\lambda-(\omega-\omega_0) t\right]}\label{eq:diffgamma}
\end{align}
}
We can formally integrate Eq. (\ref{eq:diffgamma}) and substitute into Eq. (\ref{eq:diffbeta}) to get

{\small
\begin{align}
    \dot{\beta}_j(t)=-\sum_{\lambda=L,R}\sum_{j'}\frac{\gamma_\lambda}{2\pi}\int d\omega \exp{\left[i(\omega-\omega_0)t-i\frac{\omega}{v_\lambda}x_j\right]}\nonumber\\
    \times\int_0^t dt'\beta_{j'}(t')\exp{\left[-i(\omega-\omega_0)t'+i\frac{\omega}{v_\lambda}x_{j'}\right]}
\end{align}
}
At this point, the Born-Markov approximation is usually carried out, taking $\beta_{j'}(t')=\beta_{j'}(t)$ and extending the time integration over infinity.  We take a  different approach, following the discussion in \cite{milonni1995photodetection}. Reordering phase terms and alternating integration order, we get

{\small
\begin{multline}
\dot{\beta}_j(t)=- \sum_{\lambda=L,R}\sum_{j'}\frac{\gamma_\lambda}{2\pi}\int d\omega \exp{\left[ i(\omega-\omega_0)t-i\frac{\omega}{v_\lambda}(x_j-x_{j'})\right]}\nonumber\\
\times\int_0^{t} d t' \beta_{j'}(t')\exp{\left[-i(\omega-\omega_0)t'\right]}\nonumber\\
= -\sum_{\lambda=L,R}\sum_{j'}\frac{\gamma_\lambda}{2\pi}\int_0^t dt'\beta_{j'}(t') e^{-i\omega_0(t-t')}\nonumber\\
\times\int d\omega e^{i\omega(t-t')-i\frac{\omega}{v_\lambda}(x_j-x_{j'})}
\end{multline}
}

Now we extend the frequency integration to the negative frequencies, which can be done as an approximation within the RWA, and we obtain

{\small
\begin{equation}
    \dot{\beta}_j(t) = -\sum_{\lambda=L,R}\sum_{j'}\gamma_\lambda\beta_{j'}(t-\frac{x_j-x_{j'}}{v_\lambda})e^{\left[-i\frac{\omega_0}{v_\lambda}(x_j-x_{j'})\right]}
\end{equation}
}
Here, we distinguish the retarded/advanced time $\tau_\lambda^{i,j}= \frac{x_j-x_{j'}}{v_\lambda}$. Our integration interval for $t'$ and the condition $t\geq 0$ impose the Heaviside functions for the left and right running fields, such that for the right running mode, we have $\Theta_{R}=\Theta(x_j-x_{j'})$, and for the left one, $\Theta_{L}=\Theta(x_{j'}-x_{j})$. This allows us to write

{\small
\begin{equation}
     \dot{\beta}_j(t) = -\sum_{\lambda=L,R}\sum_{j'}\gamma_\lambda\beta_{j'}\left(t-\frac{x_j-x_{j'}}{v_\lambda}\right)e^{\left[-i\frac{\omega_0}{v_\lambda}(x_j-x_{j'})\right]}\Theta_\lambda,
\end{equation}
}
from which, considering a symmetric waveguide, $\gamma_R=\gamma_L=\gamma$, $v_L=-v_R=-v$, we obtain Eq. (3) in the main text.

{\small
\begin{equation}\label{eq:mainresult}
    \dot{\beta}_j(t) = -\gamma\sum_{j'}e^{-i\omega_0\tau_{j,j'}}\beta_{j'}\left(t-\tau_{j,j'}\right)\Theta\left(t-\tau_{j,j'}\right)
\end{equation}
}

\subsection{Excitation retention in the Markovian regime}
\label{sec:Appendix2}

In the Markovian regime (i.e., $\tau=0$), one recovers the previously known results. As in ref. \cite{PhysRevA.77.043833}, the decay rates are determined by the eigenvalues of the decay matrix, each having an eigenvector. In the Bragg condition, one recovers a single superradiant eigenvector and $N-1$ dark eigenvectors. Because we use an initial condition in the form $(0,0,\dots,0,1,0,\dots,0)$, we want to express this vector as a linear sum of the decay matrix eigenvectors.\\

In the Schrödinger picture, the general solution of the system has the form
\small{\begin{multline}\label{eq:generealsolvmarkov}
|\psi(t)\rangle = C_1 e^{-i\lambda_1 t} |\lambda_1\rangle 
+ C_2 e^{-i\lambda_2 t} |\lambda_2\rangle 
+ \cdots 
+ C_N e^{-i\lambda_N t} |\lambda_N\rangle \\
+ \sum_k \alpha_k(t) |g_1 g_2 \ldots g_N\, k_\lambda\rangle,
\end{multline}}

where $\lambda_j$ is the \textit{j}th eigenvalue with the corresponding eigenvector $|\lambda_j\rangle $. At $t=0$,
\begin{equation}\label{eq:initialcond}
|\psi(0)\rangle = C_1 |\lambda_1\rangle + C_2 |\lambda_2\rangle + \ldots + C_N |\lambda_N\rangle ,
\end{equation}

In general, the system may be prepared in a state that is not an eigenstate.
\begin{equation}
|\psi(0)\rangle = \sum_{j=1}^N \beta_j(0) |g_1 \ldots e_j \ldots g_N\rangle .
\end{equation}

Replacing this in Eq. (\ref{eq:initialcond}), in matrix form, we can express this as
\small{\[
\begin{pmatrix}
\beta_1(0) \\
\vdots \\
\beta_N(0)
\end{pmatrix}
=
C_1 
\begin{pmatrix}
\lambda_{1,1} \\
\lambda_{1,2} \\
\vdots \\
\lambda_{1,N}
\end{pmatrix}
+
C_2 
\begin{pmatrix}
\lambda_{2,1} \\
\lambda_{2,2} \\
\vdots \\
\lambda_{2,N}
\end{pmatrix}
+
\cdots
+
C_N 
\begin{pmatrix}
\lambda_{N,1} \\
\lambda_{N,2} \\
\vdots \\
\lambda_{N,N}
\end{pmatrix}.
\]}

We are interested in obtaining the coefficients $C_j$. More compactly:
\[
\begin{pmatrix}
\beta_1(0) \\
\vdots \\
\beta_N(0)
\end{pmatrix}
=
\begin{pmatrix}
\lambda_{1,1} & \lambda_{2,1} & \cdots & \lambda_{N,1} \\
\lambda_{1,2} & \lambda_{2,2} & \cdots & \lambda_{N,2} \\
\vdots & \vdots & \ddots & \vdots \\
\lambda_{1,N} & \lambda_{2,N} & \cdots & \lambda_{N,N}
\end{pmatrix}
\begin{pmatrix}
C_1 \\ C_2 \\ \vdots \\ C_N
\end{pmatrix}.
\]

This has the form $\bf{M}\bf{x}=\bf{b}$, which can be solved using the method preferred by the reader. Coming back to our system, we find that a matrix of suitable eigenvectors (columns) may be written as

\begin{equation}
A=
\begin{pmatrix}
1 & -\mathbf{1}^{T} \\
\mathbf{1} & I_{N-1}
\end{pmatrix},
\end{equation}

where $\mathbf{1}\in\mathbb{R}^{N-1}$ denotes a vector whose entries are all equal to one. Then, we find the coefficients $C_i$ by inverting the matrix, which can be found using the Schur complement. 

\begin{equation}
A^{-1}
=
\begin{pmatrix}
\dfrac{1}{N} & \dfrac{1}{N}\mathbf{1}^{T} \\[6pt]
-\dfrac{1}{N}\mathbf{1}
&
I_{N-1}-\dfrac{1}{N}\mathbf{1}\mathbf{1}_{N-1}
\end{pmatrix},
\end{equation}

where $\mathbf{1}\mathbf{1}_{N-1}$ is a matrix filled with ones. Considering an initial condition where there is a single excitation in a particular atom $j$, $\beta_i(0)=\delta_{i,j}$, we find that the amplitude evolves partially as the decay rate of the superradiant state plus the frozen dynamics of the dark states. In particular, the probability for the $j$-th atom to stay excited in the steady state (after the superradiant portions has already decayed) is

\begin{equation}
   |\beta_j|^2=\left(\frac{N-1}{N}\right)^2,
\end{equation}

whereas for the other atoms ($i\neq j$), the excitation probability rises to the steady state value

\begin{equation}
    |\beta_j|^2=\left(\frac{1}{N}\right)^2.
\end{equation}

\subsection{Asymptotic scaling of the localization length}

We derive the asymptotic scaling of the localization length directly from the delay equation introduced in the main text. Starting from Eq.~(3),
\begin{equation}
    \dot{\beta}_j(t) = -\gamma\sum_{j'}e^{-i\omega_0\tau_{jj'}}\beta_{j'}(t-\tau_{jj'})\Theta(t-\tau_{jj'}).
    \label{eq:delay}
\end{equation}
We seek a stationary solutions in the long-time limit of the form
\begin{equation}
    \beta_j(t)=\phi_j e^{-iEt},
    \label{eq:stationary}
\end{equation}
where $\phi_j$ describes the spatial profile of the excitation, and $E$ is the eigenenergy, with $\hbar=1$.

Substituting Eq.~(\ref{eq:stationary}) into Eq.~(\ref{eq:delay}) gives
\begin{equation}
    iE\phi_j = \gamma\sum_{j'}e^{i(E-\omega_0)\tau_{jj'}}\phi_{j'}.
    \label{eq:eigenvalue}
\end{equation}

For a periodic array with lattice spacing $d$, $ x_j=jd$ such that $ \tau_{jj'} = \frac{|j-j'|d}{v}=|j-j'|\tau$. We assume a distance between emitters such that they satisfy the de Bragg condition, $\omega_0 \tau_{j,j'}=2\pi$. Defining $m=j-j'$, Eq.~(\ref{eq:eigenvalue}) becomes
\begin{equation}
    iE\phi_j = \gamma\sum_{m}e^{iE |m|\tau}\phi_{j-m}.
\end{equation}

Since the kernel depends only on the relative coordinate, the system is translationally invariant, and its eigenstates are Bloch waves $\phi_j=e^{ik_jd}$, where the wavenumber $k$ can take imaginary values, leading to spatial localizations. Substituting this expression yields
\begin{equation}
    iE = \gamma\sum_{m=-\infty}^{\infty}e^{iE |m|\tau}e^{-ikmd}.
    \label{eq:disp0}
\end{equation}

Introducing the dimensionless parameter $\eta=-iE\tau$, the dispersion relation can be written as
\begin{equation}
    iE = \gamma\sum_{m=-\infty}^{\infty}e^{-\eta|m|}e^{-ikmd}.
\end{equation}

The geometric series is readily evaluated,
\begin{equation}
    \sum_{m=-\infty}^{\infty}e^{-\eta|m|}e^{-ikmd} = \frac{1-e^{-2\eta}}{1-2e^{-\eta}\cos(kd)+e^{-2\eta}},\nonumber
\end{equation}
giving
\begin{equation}
    iE = \gamma\frac{1-e^{-2\eta}}{1-2e^{-\eta}\cos(kd)+e^{-2\eta}}.
    \label{eq:dispersion1}
\end{equation}

The localization length follows from the spatial distribution $\phi_j$. Only considering the evanescent component of $k$ ($i\kappa$), which determines the inverse localization length through $ \xi\propto\kappa^{-1}$. Using $\cos(i\kappa d)=\cosh(\kappa d)$, Eq.~(\ref{eq:dispersion1}) becomes
\begin{equation}
    iE=\gamma\frac{1-e^{-2\eta}}{1-2e^{-\eta}\cosh(\kappa d)+e^{-2\eta}}.
\end{equation}

Finally, in the short delay limit,
\begin{equation*}
    |\eta|\ll1,
    \qquad
    \kappa d\ll1,
\end{equation*}
the hyperbolic functions may be expanded to leading order, which yields
\begin{equation*}
    \kappa d\simeq\sqrt{2\gamma\tau}.
\end{equation*}

Therefore, the localization length in units of lattice site, matching the definition of Eq. (6) in the main text, scales as

\begin{equation}
    \xi \propto (\kappa d)^{-1}\propto \left(\gamma\tau\right)^{-1/2}
\end{equation}
demonstrating the power-law dependence of the asymptotic behavior.


\end{document}